	\documentclass[prl,twocolumn,superscriptaddress]{revtex4}
	\usepackage{amsmath}
	\usepackage[dvips]{graphicx}
	\usepackage{color}
	\usepackage{bm}

	\def \be {\begin{equation}}
	\def \ee {\end{equation}}

\begin{document}

	\title{Classical Propagation of Light in Spatio-Temporal Periodic Media}
	\author{B. S. Alexandrov} 
	\affiliation{Los Alamos National Laboratory, Los Alamos, New Mexico 87544}
	\author{K.\O. Rasmussen}
	\author{A.T. Findikoglu}
	\author{A.R. Bishop}
\affiliation{Los Alamos National Laboratory, Los Alamos, New Mexico 87544}
\author{I. Z. Kostadinov}
\affiliation {Ohio State University, Columbus,Ohio 43210l}
\date{\today }


\begin{abstract}
We analyze the propagation of electromagnetic waves in media where the dielectric constants undergo rapid temporal periodic modulation. 
Both spatially homogeneous and periodic media are studied. Fast periodic {\it temporal} modulation of the dielectric constant of a homogeneous 
medium leads to existence of photonic band-gap like phenomena. In the presence of both spatial and temporal periodicity the electromagnetic 
spectrum is described in a four-dimensional  cube, defining an effective Brillouin zone. In the case of incommensurability between space and time periodicities, 
completely dispersed point spectra exist. 
\end{abstract}
\maketitle

The advent of materials whose  electric permittivity ${\varepsilon}$ is periodically modulated at the nanometer length scale has introduced 
the important concept of photonic band-gap structures. This has in turn made traditional solid-state concepts like 
reciprocal space, Brillioun zones, dispersion relations, Bloch wave functions, etc. 
directly applicable to the field of classical electromagnetic wave propagation. 
During the last decades the attractive possible applications of such photonic band gap materials have driven 
intense theoretical and experimental studies of the propagation of electromagnetic waves in spatially periodic and
disordered dielectric structures  \cite{John_1984,Anderson_1985,Yablonovitch_1987,John_1987,Ho_1990,Leung_1990,Photonic_1994}. This
 has provided the field of photonics with a  wide range of new applications,
mostly related to guided light modes \cite{Vlasov_2005}. Recently   a magnetic photonic crystal was made \cite{Linden_2006} by periodically modulating the magnetic permeability${\mu}$ of  specially constructed material. 

Although there have also
been tentative studies  of effects arising from 
 low frequency \cite{Imai_1992}, and specific  cases  \cite{Ivanov_2001} of {\it temporal} modulation of the dielectric constant,  this possibility  has 
been largely ignored. However, as we show here, essentially all fruitful concepts from the now 
mature field of photonic band-gap materials can be applied in the case of 
fast {\it{temporal}} modulation of a material's dielectric  response. Further, the combination of 
spatial and temporal modulation of the dielectric response introduces intriguing new concepts to the field 
of classic electromagnetic wave propagation. The results are also valid for periodical modulation of the magnetic permeability ${\mu}$ of  spatially homogeneous or periodic magnetic media. 

First, it is straightforward to realize that temporal modulation leads to photonic band structures similar to spatial modulation:
In a material with a time dependent dielectric response $\varepsilon(t)$, the electromagnetic waves, $V(x,t)$, are
described by the wave equation:
\begin{equation}\label{EQ:1+1}
{\frac{\partial ^2V(x,t)}{{\partial x^2}}}={\frac{\varepsilon 
_0}{{c^2}}}{%
\frac{\partial ^2\varepsilon (t)V(x,t)}{{\partial t^2}}},
\end{equation}
where $\varepsilon_0$ is the vacuum dielectric constant  and $c$ the speed of light in vacuum.
By simple separation of variables $V(x,t)=\sum_k v_k(x)u_k(t)$,  the solution of this equation may be expressed in the form:

\begin{equation}
V(x,t)=\sum_k \varepsilon(t) u_k(t) e^{ikx} = \sum_k U_k(t)e^{ikx}.
\end{equation}
After introducing the dimensionless variables, $x\rightarrow
x/L$ and $t\rightarrow t/\tau$, the equation for $U_k(t)$ becomes:

\begin{equation} \label{EQ:FB}
\frac{d^2U_k(t)}{dt^2}+k^2s^2f(t)U_k(t)=0,
\end{equation}
where $s=\frac{c\tau }{L\sqrt{\varepsilon _0}}$ and $f(t)=\varepsilon ^{-1}(t)$.
With $\varepsilon(t)$ periodic so that $f(t)$ is periodic, the solutions will be of the Floquet-Bloch \cite{Shirley_1965} type 
\begin{equation}
U_k(t)=\exp \left (-i\omega t \right ) U_k^\omega (t),
\end{equation}
which is the central property for photonic band-gap theory.

By casting the problem in this form, it is reduced to determining the 
dispersion relation $k=k(\omega )$.  As  is the case for  traditional photonic crystals, the dispersion relation will contain the frequency regime where electromagnetic wave propagation is possible, as well as the "photonic gaps" where the propagation of electromagnetic waves is damped (or amplified) \cite{Reasons}.  
\begin{figure}[h]
   \includegraphics[width=\columnwidth]{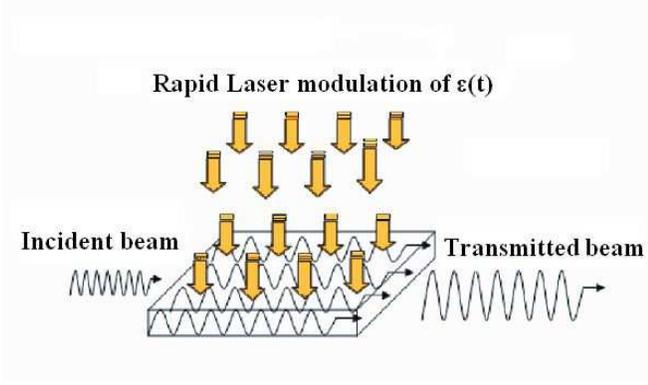}
   \caption{Illustration of concept and structure}
\label{fig:dil1}
\end{figure}
A possible realization of temporal dielectric modulation is to apply standing waves with time dependent intensity, in the form of one, two or three perpendicular laser beams, on a dielectric slab, as
illustrated in Fig. \ref{fig:dil1}. This concept is widely applied to create optical lattices within Bose-Einstein condensates \cite{Burger_2001} . Applying the beams on a spatially homogeneous or multi-dimensionally modulated 
dielectric slabs, a variety of space-time dielectric structures can be generated. 
\begin{figure}[h]
   \includegraphics[width=\columnwidth]{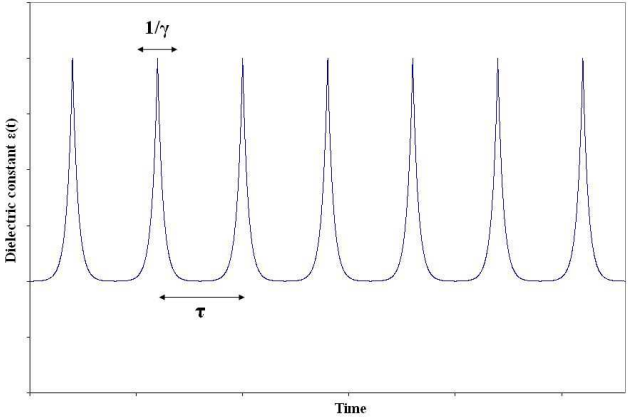}
   \caption{Illustration of the modulation of the dielectric response}
\label{fig:dil}
\end{figure} 

\begin{figure}[h]
  \includegraphics[width=\columnwidth]{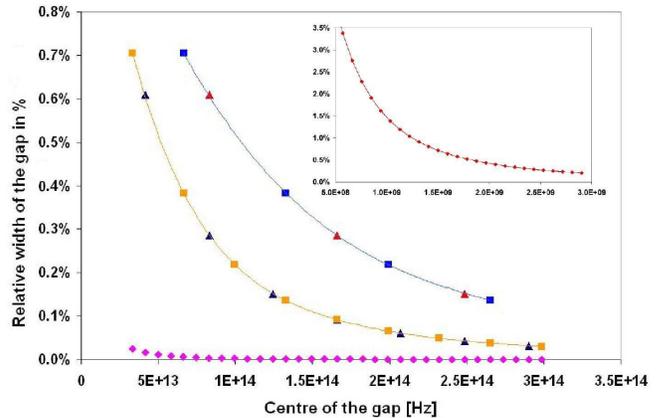}
   \caption{Width of the band-gaps as a percent of the mid-gap frequency for three frequencies $\tau
=10^{-13}$ sec - the lower curve, $\tau
=10^{-14}$ sec - the middle position curve, and $\tau
=5$x$10^{-15}$ sec - the upper curve. Triangles  correspond to $\gamma =5$, and squares  to $\gamma = 3 $.
The inset  presents  band-gaps for microwave region.  An  accessible \cite{Alp_2006}  experimental setup: $\tau
=10^{-8}$,sec  $\gamma =10$, ${\varepsilon_{temp}}=0.2{\varepsilon_0},{\varepsilon_0}=300$ was simulated.} 
\label{fig:2}
\end{figure}

First, we study the simple case of periodic dielectric modulation of a linear and homogeneous medium
by modeling it(see Fig. \ref{fig:dil})  as a series of
pulses characterized by the two times $\tau $ and $\frac{1}{\gamma}$.
The parameter $\frac{1}{\gamma}$ is the duration of a single pulse of the
dielectric function of the material $\varepsilon (t)$, and $\tau $
is the period of repetition of the pulses.
If the standing wave laser
beam consists of ultrashort pulses, 
a single pulse duration  $\varepsilon (t)$ in the range  $\gamma=10^{-12}$--$10^{-15}$sec can be realized. Similarly,
a pulse separation $c\tau$ comparable to the
wavelength of the light in the medium is realistic. Specifically, we  study the propagating waves in a $1+1$ space-time scalar wave equation (\ref{EQ:1+1}) by assuming the dielectric constant to be of the form:
\begin{eqnarray}
\varepsilon (t)&=&1+\sum_{m=1}^Ng(t-m\tau ) \nonumber \\ &=& 1+\sum_{m=1}^N\frac{\epsilon_{temp}\gamma \tau}{2} \exp \left (-\gamma | t- m \tau| \right).
\end{eqnarray}
The specific choice (see Fig. \ref{fig:dil} ) of the function $g(t)$ is motivated by its convenient form for
analytical treatment: For narrow pulse widths ($\gamma \rightarrow \infty$), 
$g(t)$ reduces to a $\delta $-function and the model is 
then similar to the Kronig-Penney model. 
Applying this expression in Eq. (\ref{EQ:FB}) 
for the Fourier amplitude $\bar{U}_k(\omega )$:
\begin{equation}\label{EQ:Fourier}
\left( \omega ^2-k^2s^2\right) \bar{U}_k(\omega )=\sum_{m 
=1}^N{\frac{{%
\omega ^2\varepsilon ^t\gamma ^2e^{-i\omega m}}}{{\left( 
\gamma
^2+\omega ^2\right) }}}U_k(m),
\end{equation}
where we have used the approximation 
\begin{equation}\nonumber
\int dte^{-\gamma |t-n|}U_k(t)f(t)\approx U_k(n)\int dte^{-\gamma
|t-n|}f(t),
\end{equation}
which amounts to assuming the wave function $U_k(t)$ to be constant 
during the pulse duration $1/\gamma$ (see Fig. \ref{fig:dil}).
For the discrete set of amplitudes $\bar{U}_k(\omega )$, when we
integrate both sides of Eq. \ref{EQ:Fourier}, and employ the relationship
\begin{equation} \nonumber
U_k(m)\equiv U_k(t=m)\equiv
\int\limits_{-\infty }^\infty {\frac{d\omega }{{2\pi }}}e^{i\omega
m}\bar{U}_k(\omega ),
\end{equation}
we obtain the matrix equation:
\begin{equation}
U_k(m)=\sum_{m=1}^N\Lambda _{n,m}U_{k_n}(m 
),
\end{equation}
\begin{equation}
\Lambda _{n,m}={\frac{\varepsilon_{temp}\gamma ^2}{{2\pi }}}%
\int\limits_{-\infty }^\infty {\frac{{e^{i\omega (n-m)}\omega 
^2}}{%
{(\omega ^2-k^2s^2)(\omega ^2+\gamma ^2)}}}d\omega.
\end{equation}
Without further approximations we find after some manipulation (see supplement for details) 
that the dispersion relation of this system can be expressed as:
\begin{equation}\label{EQ:time_crystal}
\cos \omega =h_{temp}^2-\sqrt{(h_{temp}^1)^2-4h_{temp}^0},
\end{equation}
where
\begin{equation}\nonumber
h_{temp}^1=\frac{1}{2}\left[ \cosh \gamma +\cos ks+\eta_{temp}(\cosh 
\gamma +\frac{ks}{\gamma }\sin ks)\right],
\end{equation}
\begin{equation}\nonumber
h_{temp}^0=\cosh \gamma \cos ks+\eta_{temp}(\sinh \gamma \cos 
ks+\frac{ks}{\gamma}\cosh \gamma \sin ks),
\end{equation}
and
\begin{equation}\nonumber
\eta_{temp}=\frac{\gamma^3}{2(\gamma^2+k^2s^2)}\epsilon_{temp}.
\end{equation}
\begin{figure}[h]
 \includegraphics[width=\columnwidth]{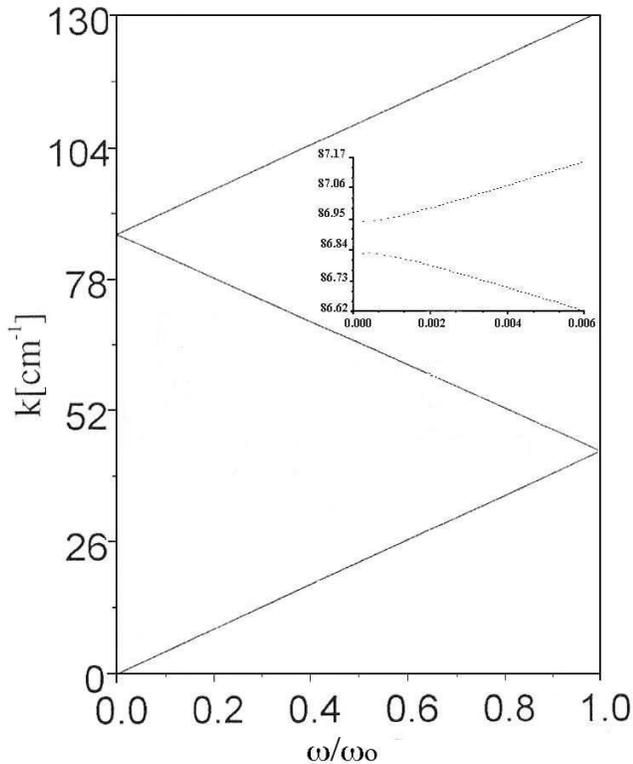}
   \caption{Spectrum of a ''time photonic crystal'' with $\tau
=4$x$10^{-12}$ sec; $\gamma =4$; $\varepsilon _0=11$;
$\varepsilon_t=0.01$. The ''band-gap'' is displayed in the upper
inset.} 
\label{fig:1}
\end{figure}
In Fig. \ref{fig:1} we illustrate this dispersion relation, Eq. \ref{EQ:time_crystal}.
The width of the gap is very small because of the realistic 
\cite{Yablonovitch_1993} smallness of the modulation $\frac{\varepsilon_{temp}}{\varepsilon_0}=0.01$ of the dielectric constant. One can easily see that the widths of the gaps are proportional to $\varepsilon_{temp}$.
Figure \ref{fig:2} shows the width of the band-gap as a function of   $\tau$ and  $\gamma$. This width  is much larger for high frequencies but is relatively insensitive to $\gamma$. 
At this point it is instructive to note that for a traditional one dimensional photonic crystal, similarly defined by the dielectric constant
\begin{eqnarray}
\varepsilon (x)= 1+\sum_{m=1}^N\frac{\epsilon_{spat}\rho L_x}{2} \exp \left (-\rho | x- m L_x| \right),
\end{eqnarray}
we obtain the dispersion relation by simply interchanging $k$ with $\omega$ and $ks$ with $\frac{\omega}{s}$.

Turning now to the more general question of multidimensional periodicity, we
assume the dielectric constant $\varepsilon (r,t)$ to be a
periodic function of both space and time, and solve the Maxwell equations for the electric field $\vec
E(r,t)$. Again we have the Bloch type solutions with
respect to space and time variables and so we  introduce
Bloch-Floquet  parameters.
The final dispersion relation will then relate
${\bf k}$ and $\omega $.  This means that in the case of combined spatial and temporal
periodicity the electromagnetic excitations spectrum is
described by separated points represented by $(k,\omega )$,
all belonging to an equivalent four-dimensional cube defining the four-dimensional
Brillouin zone. For simplicity, we  consider the dielectric constant $\varepsilon
({\bf r},t)$ to be of the type $\varepsilon ({\bf r},t)=\varepsilon ({\bf r})\varepsilon (t)$, i.e. we  assume
independent spatial and temporal modulations of the dielectric media.
The general solution of the wave equation is then
\begin{equation}
\vec {\bf E(r},t)=\sum_n\vec {\bf V}_n({\bf r})U_n(t),
\end{equation}
where for $\vec {\bf V}_n({\bf r})$ and $U_n(t)$ we have
\begin{equation}\label{E1}
{\frac{d^2}{{dt^2}}}\varepsilon (t)U_n(t)=-\nu ^2s^2U_n(t)
\end{equation}
and
\begin{equation}\label{E2}
{\nabla \times \nabla \times \vec {\bf V}}_n({\bf r})= -\nu
^2\varepsilon ({\bf r})\vec {\bf V}_n({\bf r})
\end{equation}
with
\begin{equation}\nonumber
\nabla \cdot \varepsilon ({\bf r})\vec {\bf V}_n({\bf r})=0
\end{equation}
and  $s=\frac{c\tau }{L_x\sqrt{\varepsilon _0}}\equiv \frac{L_t}{L_x}$ 
and $\nu $ being constants. 
\begin{figure}[h]
 \includegraphics[width=\columnwidth]{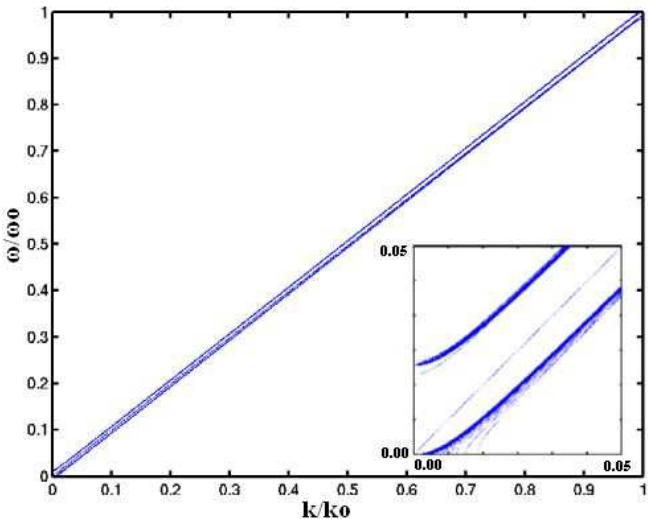}
   \caption{Dispersion relation of the temporally and spatially modulated medium with  with 
$ \tau=3$x$10^{-13}$ sec; $\gamma =3$; $\varepsilon _0=11$;$L_x=3$x$10^{-5}$ m ; $ \rho =3$ ;$\varepsilon_x=0.01$;
$\varepsilon_t=0.01$ s=1.  The inset shows a blow-up illustrating the underlying complex structure of the dispersion relation.}
\label{fig:3}
\end{figure}
\begin{figure}[h]
   \includegraphics[width=\columnwidth]{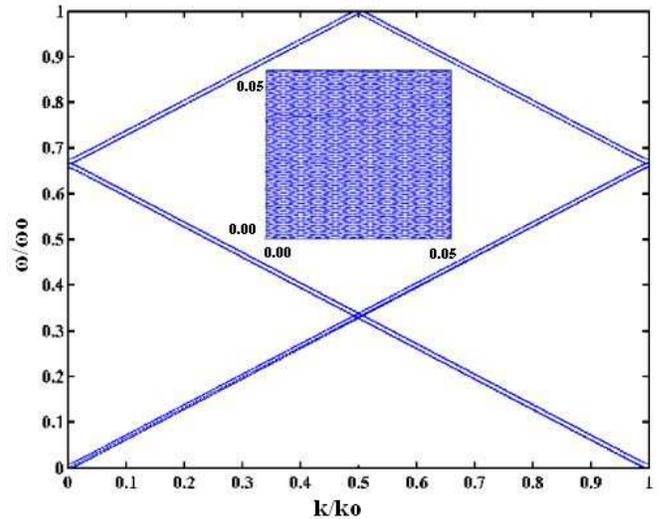}
   \caption{Spectrum of the ''space-time photonic crystal'' with $\tau
=3$x$10^{-13}$ sec; $\gamma =3$; $\varepsilon _0=11$;$\varepsilon_x=0.01$;
$\varepsilon_t=0.01$ $s=\frac{2}{3}$. In the inset we represent a point spectrum - the incommensurability case of two basic lengths, where $s=\sqrt{2}$ and $\tau
=3$x$10^{-13}$ sec.} 
\label{fig:4}
\end{figure}
In order to simplify the problem, we consider
the wave equation for the scalar wave
amplitude $V(x,t)$. In this  case of periodicity in one space dimension and in time, we have to take into account that 
the separation constant $-\nu^2$ is the same in both equations (\ref{E1}) and (\ref{E2}). 

We can now establish the dispersion relation between the normalized frequency $\omega$ 
and the normalized wave-number $k$. For simplicity we first examine the case of $s=1$, where the temporal and 
spatial periodicities are equal, $L_t=L_x$. We find that the dispersion  of such a media is dominated by a vacuum -like dispersion relation i.e. $\omega \simeq sk$. The spectrum in this  case is given mostly by the diagonal of the unit square, resembling in this sense a homogeneous medium. The actual 
dispersion  depicted in Fig. \ref{fig:3} shows this overall behavior but, as  can be clearly seen in the inset, the underlying structure is very complex. In fact the dispersion relation is a dense set of discrete points clustering around two separate branches.
From this case we can easily construct the overall structure of the dispersion relation resulting from 
rational values of $s\neq1$.  For an integer values of $s>1$  the relation $\omega \simeq sk$ results in 
values of $\omega>1$  and so this part of the dispersive branch is 
folded back into the standard zone.
Similarly, for integer values of $s^{-1}> 1$ the roles of $\omega$ and $k$ are
reversed and the $k$ values have to be folded back into the standard zone. 
Therefore, for rational values of $s =M/N$, where 
$M$ and $N$ are integers leads to the repetition of $M$ branches along the $k$
axis and $N$ branches along the $\omega$ axis. This is illustrated  in 
Fig. \ref{fig:4}, where the $s=2/3$ case is displayed. When $s$ is not a rational number, every single line determining the dispersion relations has a definite width. This  means that in these cases any $\omega$ 
corresponds to an infinite number of $k$ values, and {\em vice-versa}.
{\it{This is completely new situation, which has no analog in photonic or electronic band structures}}. The
reason for this broadening  is the existence of small but finite
band-gaps. The folding of the dispersion curves leads to a slight displacement 
of the equivalent curves. The
largest displacements appear for the intervals of $\nu$ where
gaps are large. Then, at the larger values of $\nu$, the
curves become closer  and when $\nu$ tends to infinity they
approach the corresponding straight line branches. This means that, when $s$ is not a rational number,  incommensurability  between the spatial and temporal periodicities exists. In this case the dispersion relation consists of separated points and it is intriguingly complex, as is illustrated for $s=\sqrt{2}$ in the inset of  Fig. \ref{fig:4}. Finally, we emphasize that all these effects exist when the time variation of $\varepsilon(t)$ is rapid. When $\varepsilon(t)$ is a slowly varying function of  time compared to the amplitude $U_{k}(t)$, we can simply ignore its time dependence \cite{Slow}.

In summary,  we have studied the effects of 
time variation of the dielectric constant of different electromagnetic media, which
leads to the existence of band-gap like phenomena. The physical reason for these effects is the
necessity of synchronization of the phases of the propagating
waves with the external time periodicity of the medium. In  cases of
simultaneous space and time periodicity, we found an
electromagnetic wave spectrum essentially described in an
equivalent two-dimensional, three-dimensional or even four-dimensional cube defining the Brillouin zone.
The folding of the dispersion relation 
when $s=M/N$, with $M$ and $N$ integer, and broadening of
dispersion curves have been demonstrated. When $s \neq M/N$ the effects of incommensurability
of the two internal lengths ( space $L_x$ and time $L_t = {\frac{c\tau 
}{{\sqrt{\varepsilon_o}}}}$ )exist, and point like spectra are
exhibited. Modulation of the photonic band structure is suggested.

One of us, B.S.A. is  grateful to Prof. E.Yablonovitch and Prof. M. Balkanski for the useful
discussions. I.Z.K. and B.S.A.  would like to acknowledge the assistance of Dr.V.Popov and Prof.M.Mateev.
Work at Los Alamos National Laboratory is performed under the auspices of the US DoE.


\begin{thebibliography}{99}

\bibitem{John_1984}  S. John, Phys. Rev. Lett. {\bf 53}, 2169 (1984)
\bibitem {Anderson_1985} P. W. Anderson, Phil. Mag. {\bf B 52}, 5050 (1985)
\bibitem{Yablonovitch_1987} E. Yablonovitch , Phys. Rev. Letters , 58 , 2059 (1987)
\bibitem {John_1987} S. John, Phys.Rev.Lett. {\bf 58}, 2486 (1987)
\bibitem {Ho_1990}  K. M. Ho, C. T. Chan, and C. M. Soucoulis, Phys. Rev. Lett. {\bf 65},3152 (1990)
\bibitem {Leung_1990}  K. M. Leung and Y. F. Lin, Phys. Rev. Lett. {\bf 65}, 2636 (1990)
\bibitem {Photonic_1994} Photonic Band Structure, Special Issue, J. of Mod. Optics,V.{\bf41}, No.2 (1994)
\bibitem {Vlasov_2005}Yurii A. Vlasov, Martin O'Boyle, Hendrik F. Hamann and  Sharee J. McNab, Nature, Vol 438, 65,(2005)
\bibitem {Linden_2006} S. Linden, M. Decker and M. Wegener Phys. Rev. Lett. {\bf} 97, 083902 (2006) 
\bibitem{Imai_1992} Masaaki Imai, Takashi Yano, Kazushi Motoi, Akira Odajima,IEEE Vo.{\bf28}, No.9, 1901
\bibitem{Ivanov_2001} Ivanov A. L.  and Littlewood P. B.  Phys. Rev. Lett. 87 136403 (2001) Ivanov A. L. and Littlewood P. B.  Patent GB 0121448.5.(2001).
 \bibitem{Shirley_1965} Shirley J.H. Physical Review Vol. 138, 4B, p. B979 (1965).
\bibitem{Reasons} The physical reason for the band-gap like phenomena is the necessity of synchronization of the phases of the propagating waves with the external time periodicity of the medium. The additional energy for amplification will be proportional to the ${\frac{\partial P}{{\partial t}}}$, where $P$ is the induced polarization.  
\bibitem{Burger_2001} S. Burger, F. S. Cataliotti, C. Fort, F. Minardi, M. Inguscio, M. L. Chiofalo, and M. P. Tosi. Phys. Rev. Lett., 86:4447, 2001.
\bibitem{Alexandrov_1993} B.S. Alexandrov and I.Z. Kostadinov, ``Time Photonic Crystal'', presented on  ``Confined Electrons and Photons: New Physics and Devices'',  NATO/ASI,  Erice (Italy),  (1993) 
\bibitem{John_1990}  S. John and J. Wang, Phys. Rev. Lett. {\bf 64}, 2418 (1990)
\bibitem{Yablonovitch_1993}  E. Yablonovitch, private communication.
\bibitem{Alp_2006}   An experimental test of these ideas could  be made at microwave 
frequencies. The response of most nonlinear dielectric materials is much 
stronger at microwave frequencies than at optical frequencies.  For example, 
 in $Sr_{1-x}Ba_{x}TiO_{3}$ compounds, it is possible to achieve $ \varepsilon_{t}=0.2 \varepsilon_{0} $ with 
$\varepsilon_{0} =300$ for an applied electric field of about $10^6 V/m$ at frequencies up to 
several tens of GHz. In addition, at microwave frequencies, network 
analysis equipment and techniques could be employed to achieve high 
spectral resolution, good signal sensitivity, and large dynamic range.
\bibitem{Andre_2002} A. Andre, M. D. Lukin,  Phys. Rev. Lett. {\bf 89},143602, (2002)
\bibitem{Slow} In this case one can, as a first approximation, substitute $\varepsilon=\varepsilon(t)$ directly in the condition for the effective "Bragg resonance" and  "Mie-resonance". As a result we will have a photonic crystal with filling fraction $f=1/(2\sqrt{\varepsilon(t)})$ breathing in time, i.e we  have a time modulation of the photonic band structure.

\end{thebibliography}
\end{document}